\documentstyle[12pt]{article}

\parskip=\baselineskip
\begin{document}
\begin{center}
\vspace {40mm}
{\bf {\Large EUCLIDEAN FIELD THEORY AND SINGULAR CLASSICAL FIELD
CONFIGURATIONS}}\\
A. Shurgaia \\ {\it Department of Theoretical Physics \\ Mathematical Institute
of Georgian Academy of Sciences\\ Tbilisi Republic of Georgia}.\footnote
{E-mail address: avsh@imath.acnet.ge}\\
\vspace{10mm}
{\bf Abstract }\\
\end{center}
Euclidean field theory on four dimensional sphere is suggested for the study
of high energy multiparticle production. The singular classical field
configurations are found in scalar $\phi^4$ and SU(2) gauge theories and
the cross section of 2$\rightarrow n$  is calculated. It is shown, that
the cross section has a maximum at the energy compared to the sphaleron
mass.

\newpage

One of the most interest topics of the last years is that of
high energy multiparticle production, in which the final many-particle
state can be described as a semiclassical one and when the classical solutions
play an important role (examples of such processes are the electroweak
processes, accompanied by baryon number violation[1-6], multijet production
in strong interactions[7,8]). By investigating these questions we come to
the calculational problem of dealing with $2\rightarrow n$ processes with the
final state considered semiclassically, but the initial state as a quantum 
state.  Describing the final state as a semiclassical with definite energy and
going back in time to the initial two-particle state, one violates the energy
conservation low. This means that we have to consider the transition between
states with the deferent energies. A way to circumvent this difficulty
is suggested in the consideration of the singular classical trajectories in the
imaginary (Euclidean) time, which were introduced by Landau 60 years ago in
the calculation of the transition probability between the low and high
energy quantum mechanical states[9]. This approach has been generalized to
the quantum field theory by Iordanskji and Pitaevskji[10]. On the basis of
their approach S.Yu. Khlebnikov[11] has suggested the use of singular
Euclidean solutions for study of the above mentioned problem. The result of
Landau has been used by M.Voloshin[12] in four dimensional $\phi^4$ theory
by considering spatially constant fields. Recently D.Diakonov and
V.Petrov[13] developed this approach for the double-well potential and applied
to the Yang-Mills theory. They obtained some approximate singular solutions
in pure Yang-Mills and elektroweek theories. At low energies their results
coincide with those of instanton-induced multiparticle production cross
section, but at higher energies they yield exponentially decreasing behavior
of cross section in the running gauge coupling constant. The cross
section has a maximum at the energy defined by sphaleron mass.

1. EUCLIDEAN FIELD THEORY APPROACH ON $4^4$.

We propose below the study of above mentioned problem by the method
based on a field theoretical approach, which has been developed in [14]. In
particular it has been suggested an Euclidean field theory on sphere $S^4$
and has been shown, that the system evolves along the
radius of $S^4$. The dilatation operator is the evolution operator. This
approach is convenient especially in scale invariant theories, in which the
dilatation operator can be diagonalized. We recapitulate briefly some
salient features of this approach.
Consider for simplicity the real
scalar massless field $\phi(x)$ in Euclidean space. Following[14] we
introduce the spherical coordinates:
\begin{eqnarray}
x_\mu=r{\bf \alpha}_\mu(\vartheta,\varphi,\psi),
\end{eqnarray}
where ${\bf \alpha}_\mu(\theta, \varphi,\psi)$ is a unit
vector. Defining new field $\chi (r,{\bf \alpha})=r\phi (r,{\bf \alpha})$ the
action is written
\begin{eqnarray}
S=\frac{1}{2}\int_0^\infty\frac{dr}{r}\int d{\bf
\alpha}\Bigl\{\bigl(r\frac{\partial \chi}{\partial r}\bigr)^2\Bigr.
\Bigl.+\chi(\hat L^2+1)\chi\Bigr\},
\end{eqnarray}
where $\hat L^2$ is a square of angular momentum.
For field $\chi (r,{\bf \alpha})$ the following commutation relations hold:
\begin{eqnarray}
[\chi (r,{\bf \alpha}),\chi (r,{\bf \alpha}')]=[\dot \chi (r,{\bf \alpha}),
\dot \chi (r,{\bf \alpha}')]=0, \qquad
[\dot \chi (r,{\bf \alpha}),\chi (r,{\bf \alpha}')]=-{\delta^3}({\bf \alpha}
-{\bf \alpha}'),
\end{eqnarray}
Here we introduced the notation: $\dot \chi (r,{\bf \alpha}%
)=\partial \chi (r,{\bf \alpha}) /\partial (\ln r)$ and the $\delta$ function is defined on
the sphere:
\begin{eqnarray}
\int d{\bf \alpha}_1f({\alpha}_1)\delta^3 ({\bf \alpha}_1-{\bf \alpha}_2)=
f({\bf \alpha}_2).
\end{eqnarray}
The nonvanishing commutator becomes usefull, if we introduce
the proper time $\tau =-i\ln r$. Really
\begin{eqnarray}
[\dot \chi (\tau ,{\bf \alpha}),\chi (\tau ,{\bf \alpha}')]=-i\delta^3 ({\bf \alpha}
-{\bf \alpha}'),
\end{eqnarray}
which is formally similar to the equal-time commutator of
conventional field theory. The equation of motion for field $\chi(r,{\bf %
\alpha})$ is
\begin{equation}
\frac{\partial^2\chi (r,{\bf \alpha})}{\partial(\ln r)^2}-(\hat L^2+1)\chi (r,{\bf \alpha})=0
\end{equation}
allows the separation of the variables. The eigenfunctions of the angular
operator $\hat L^2$ form a complete orthonormal set of spherical functions
$Y_{lnm}(\vartheta, \varphi,\psi)$:
\begin{equation}
\hat
L^2Y_{lnm}(\vartheta,\varphi,\psi)=l(l+2)Y_{lnm}(\vartheta,\varphi,\psi)
\end{equation}
and are given by
\begin{equation}
Y_{lnm}(\vartheta,\varphi,\psi)=N_{lnm}e^{im\varphi}\sin^n \vartheta
G^{n+1}_{l-n}(\cos \vartheta)\sin^m \psi G^{m+1/2}_{n-m}(\sin \psi),
\end{equation}
where $N_{lnm}$ is a normalization constant and $G_n^m(x)$ is a Gegenbauer
polynomial. The numbers $l,n,m,$ are integers:
\begin{equation}
l=0,1,2,...,\qquad n=0,...l, \qquad m=-n,...n.
\end{equation}
Taking into account the radial part of $\chi (\tau,{\bf \alpha})$ and using the
proper time $\tau$ one can write the following expansion for $\chi (\tau,%
{\bf \alpha})$:
\begin{eqnarray}
\chi (\tau,{\bf \alpha})=\sum_{l=0}^{\infty} \sum_{n=0}^l \sum_{m=-n}^n
 \bigl[{a^{(-)}}_{lnm}\frac{e^{-i\tau (l+1)}}{2l+2}Y^*_{lnm}({\bf \alpha})\bigr.
\bigl.+{a^{(+)}}_{lnm}\frac{e^{i\tau (l+1)}}{2l+2}Y_{lnm}({\bf \alpha})\bigr].
\end{eqnarray}
The hermiticity condition applied to $\chi (\tau,{\bf \alpha})$ shows, that
$\chi (\tau,{\bf \alpha})$ is Hermitian for real $\tau$ and in this case with
$[a^{(-)}_{lnm}]^+=a^{(+)}_{lnm}$.
Considering ${a^{(-)}}_{lnm}$ as an annihilation operator, we define the
vacuum state $\vert 0>$ as being annihilated by all ${a^{(-)}}_{lnm}$
operators:
\begin{eqnarray}
{a^{(-)}}_{lnm}\vert 0>=0.\nonumber
\end{eqnarray}
Next one can compute the vacuum expectation value of the product of two
operators. In this way  the following remarkable expression for
the propagator $D(x_1-x_2)$ is obtained: $(\Box D(x_1-x_2)=-\delta^4(x_1-x_2))$:
\begin{eqnarray}
D(x_1-x_2)=\frac{1}{4\pi^2 \vert x_1-x_2\vert}=\frac{1}{2\pi^2}\int_{-\infty}^
\infty d\varepsilon \sum_{lnm}\frac{{F^*}_{lnm}(x_1)F_{lnm}(x_2)}{\varepsilon^2+
(l+1)^2}.
\end{eqnarray}
The functions $F_{lnm}(x)=r^{i\varepsilon -1}Y_{lnm}({\bf %
\alpha})$ define the transformation between the Euclidean coordinate space and the
conjugate space $(\varepsilon lnm)$, in which the dimensionality $\varepsilon$
and the quantum numbers $l,n,m$ result from diagonalization
of their respective dilatation and angular momentum operators. In this space, in particular
for scale invariant theory the propagator is diagonal:
\begin{equation}
D(\varepsilon,l,n,m)=\frac{1}{\varepsilon^2+(l+1)^2} \nonumber
\end{equation}
and has poles at the eigenvalues of the evolution operator.

Next we consider the interaction with the external source. The original equation of
motion in Euclidean coordinates reads:
\begin{equation}
\Box \phi (x)=\eta (x).\nonumber
\end{equation}
One may expand $\chi (\tau,{\bf \alpha })$ as follows:
\begin{eqnarray}
\chi (\tau,{\bf \alpha})=\sum_{l=0}^{\infty} \sum_{n=0}^l \sum_{m=-n}^n
\bigl[{A^{(-)}}_{lnm}(\tau)\frac{e^{-i\tau (l+1)}}{2l+2}{Y^*}_{lnm}({\bf \alpha})\bigr.
\bigl.+{A^{(+)}}_{lnm}(\tau)\frac{e^{i\tau (l+1)}}{2l+2}Y_{lnm}({\bf \alpha})\bigr],
\end{eqnarray}
so that, the quantities ${A^{(}\pm )}_{lnm}(\tau)$ obey the
following equation:
\begin{eqnarray}
\frac{d{A^{(\pm)}}_{lnm}(\tau) }{d\tau}=\pm(l+1){A^{(\pm)}}_{lnm}(\tau)\pm
\frac{(-1)^{(m\pm m)/2}}{2l+2}{\eta^{\pm}}_{lnm}(\tau),
\end{eqnarray}
where $\eta _{lnm}^{\pm }(\tau)$ is a $(\varepsilon lnm)$ transform of
source function. Introducing the evolution operator one can show, that it is
defined by
\begin{eqnarray}
U(r,r_0)=R\exp \bigl\{-i\int^r_{r_0}d(\ln r') D^{int}(r')\bigr\},
\end{eqnarray}
where $R$ denotes the ordering along the radius of $S^4$, and
$D^{int}(r)$ is an interaction part of dilatation operator. The integration
is over all space bounded by the spheres with $r^2_0<r^2<r^2$. It will be
noted, that the scale invariance of the theory is not necessary requirement. In
that case dilatation operator is still an evolution operator, but can not be
diagonalized. In principle all scale-breaking terms can be treated
perturbativly.

Comparing this approach with the conventional theory constructed on $t=const$
surface we see, that the energy and 3-momentum are replaced by
dimensionality and the numbers $(lnm)$. So in considering the multiparticle
production iniciated by annihilation of two particles we describe the initial
and final states by dimensionality $\varepsilon$ instead of energy. In order
to obtain the physical on shell amplitude we have to calculate first the
two-point Green function, take its transform in $(\varepsilon lnm)$ space
(instead of Furier transform in conventional theory) and then apply the
procedure of LSZ. In this way we connect the probability of reaction with
the full Green function.
\newpage
2. LSZ FORMALISM.

We recall now briefly the LSZ reduction formalism in the context of field theoretical
approach suggested in[13]. For simplicity we consider a scalar massless
field. Let us  postulate first $|in>$ and $|out>$ states as the asymptotic states of
the interacting field in the limits $\tau \rightarrow -\infty $ and $\tau
\rightarrow \infty $ respectively. These asymptotic fields
$\chi_{in}(\tau,{\bf \alpha })$ and $\chi _{out}(\tau,{\bf \alpha })$ satisfy
free equation of motion:
\begin{eqnarray}
{\hat {\bf A}}(\tau,{\bf \alpha})\chi_{\stackrel{in}{out}}(\tau,{\bf \alpha})\equiv
\Bigl(\frac{\partial^2}{\partial\tau^2}+({\hat L^2}({\bf \alpha})+1)\Bigr)
\chi_{\stackrel{in}{out}}(\tau, {\bf \alpha})=0 \nonumber
\end{eqnarray}
We define the following ''time''-independent scalar product of two functions:
\begin{eqnarray}
<\phi_1, \phi_2> = \int d{\bf \alpha} \phi_1 \frac{\stackrel{\leftrightarrow}
{\partial}}{\partial\tau}\phi_2.
\end{eqnarray}
For fields $\chi _{\stackrel{in}{out}}(\tau,{\bf \alpha })$ the expansion (10)
holds. The creation  operator $a^{+}_{\stackrel{in}{out}}(lnm)$ is expressed
through $\chi _{\stackrel{in}{out}}(\tau {\bf \alpha })$ as follows:
\begin{eqnarray}
a^{+}_{\stackrel{in}{out}}(lnm)=-i\int d{\bf \alpha}e^{-i\varepsilon\tau}Y_{lnm}
({\bf \alpha})\frac{\stackrel{\leftrightarrow}{\partial}}{\partial\tau}
\chi_{\stackrel{in}{out}}(\tau {\bf \alpha}). \nonumber
\end{eqnarray}
Let us denote the complete set of $(\varepsilon lnm)$ through $q$ and consider the
transition amplitude $<{q'}_1,\ldots ,out|q_1,\ldots ,in>$
\begin{eqnarray}
<q'_1, \ldots,out\vert q_1, \ldots, in>=<q'_1, \ldots, out\vert a^{+}_{in}
\vert q_2,\ldots,in>=\nonumber  \\
-i\int d{\bf \alpha}e^{-i\varepsilon\tau}Y_{lnm}({\bf \alpha})
\frac{\stackrel{\leftrightarrow}{\partial}}{\partial\tau_1}
<q'_1, \ldots, out\vert \chi_{in}(\tau_1, {\bf \alpha})
\vert q_2, \ldots, in>.
\end{eqnarray}
Since the integral is independent of $\tau_1$
\begin{eqnarray}
&<q'_1, \ldots,out\vert q_1,\ldots,in>=&   \nonumber \\
&-i\lim_{\tau_1 \rightarrow -\infty}\int d{\bf \alpha}e^{-i\varepsilon\tau}
Y^{*}_{lnm}({\bf \alpha})
\frac{\stackrel{\leftrightarrow}{\partial}}{\partial\tau_1}
<q'_1,\ldots,out\vert \chi (\tau_1,{\bf \alpha})\vert q_2, \ldots,in>,&
\end{eqnarray}%
which can be reduced to
\begin{eqnarray}
&&<q'_1,\ldots,out\vert q_1,\ldots,in>=<q'_1,\ldots,out\vert a^{+}_{out}
\vert q_2,\ldots,in>+ \nonumber  \\
&&i\int d\tau_1d{\bf \alpha}e^{-i\varepsilon\tau}Y_{lnm}({\bf \alpha})
\Bigl[\frac{\partial^2}{\partial \tau^2_1}+{\hat L^2}({\bf \alpha})+1\Bigr]
<q'_1,\ldots,out\vert \chi(\tau_1, {\bf \alpha})\vert q_2,\ldots,in>.
\end{eqnarray}
The first term represents a disconnected part of amplitude. By
repeating this reduction step by step we obtain the following result (apart
from disconnected terms):
\begin{eqnarray}
& &<q'_1\ldots q'_k, out\vert q_1\ldots q_s in)= \nonumber \\
& &i^{k+s}\int d\tau_1d{\bf \alpha_1} \ldots d\tau_s d{\bf \alpha_s}
e^{(i\sum_{j=1}^{k}\varepsilon_j\tau_j- \sum_{j=1}^{s}\varepsilon_j\tau_j)}
\prod_{j=1}^{k}Y_{l_jn_jm_j}({\bf \alpha_j})\prod_{j'=1}^{s}Y^{*}_{l_{j'}n_{j'}m_{j'}}({\bf \alpha_j'})\nonumber \\
& &\times {\hat {\bf A}}(\tau_1,{\bf \alpha_1})\ldots{\hat {\bf A}}(\tau_s,{\bf \alpha_s})
<0\vert R\chi(\tau_1 {\bf \alpha_1})\ldots \chi(\tau_s {\bf \alpha_s})\vert0>.
\end{eqnarray}
The symbol $R$ expresses the fact, that the product of
operators is ordered along the radius of sphere $S^4$. It will be mentioned,
that on mass shell condition of conventional theory is replaced here by $%
\varepsilon^2\rightarrow (l+1)^2$. Having this scheme we can consider the 
asymptotic behavior
of Green function - to be more exact its $(\varepsilon lnm)$ transform - for
large $\varepsilon$
similar to the high energy behavior of the Fourier transform of the Green function,
considered in[10], and show the importance of singular trajectories.

3.  CLASSICAL SINGULAR TRAJECTORIES AND CROSS SECTION.

We shall follow to paper[13] in order to calculate semiclassically the total
cross section induced by scattering of two particles, connecting it with the help
of the optical theorem with the imaginary part of the diagonal matrix element of
the scattering amplitude \\
$<\varepsilon lnm\vert M\vert\varepsilon lnm>$:
\begin{eqnarray}\sigma(\varepsilon)\sim &&\lim_{\varepsilon^2\to(l+1)^2}Im\int d\tau_1...d\tau_4
d{\bf \alpha}_1...d{\bf \alpha}_4 \exp \{-i\varepsilon_1(\tilde\tau_1-\tilde\tau_3)-
i\varepsilon_2(\tilde\tau_2-\tilde\tau_4)\}\nonumber\\
&&\times Y^*_{lnm}({\bf \alpha}_1) Y^*_{lnm}({\bf \alpha}_2) Y_{lnm}({\bf \alpha}_3)
Y_{lnm}({\bf \alpha}_4) <{\hat {\bf A}_1} \chi(\tilde\tau_1,{\bf \alpha}_1)...
{\hat {\bf A}_4} \chi(\tilde\tau_4,{\bf \alpha}_4)>_0,
\end{eqnarray}

The expression (22), which is our starting formula, looks like the corresponding
formula of conventional field theory with replacement $t\to\tilde\tau,\;
E\to\varepsilon,\; \vec k\to (lnm)$. The requirement of mass shell condition
is replaced by $\varepsilon^2\to (l+1)^2$. According to [8],[9],[12] we
arrive at singular trajectories, parametrized by pure imaginary time
$\tilde\tau=-i\ln r=-i\tau$.

In our field theory the trajectory begins at $\tau=-\infty$ from
vacuum (we suggest that the potential $U(\chi)$ is double-well), where $%
\varepsilon=0$ and goes to the singularity at some value of $\tau=-\tau_0/2$%
. At this point the dimensionality $\varepsilon$ receives an increment and
field proceeds further with the fixed $\varepsilon$ to the first turning
point at $\tau=0$. At the turning point the field can enter in principle the
region, where $\tau$ is real ( it corresponds to the Minkowskian part of
conventional theory). But in this region the exponential is pure phase. Next
the amplitude can be squared. It means, that the trajectory has to be
replaced in opposite direction going from turning point with fixed nonzero $%
\varepsilon$ to the singularity then returning ultimately to the vacuum.
Finally one can be obtained the following result for the cross section (up
to the exponential accuracy):
\begin{equation}
\sigma (\varepsilon) =e^{-S(\varepsilon)}
\end{equation}
where $S$ is full classical action and
\begin{equation}
S=S^{{\rm I}}-S^{{\rm II}}-S^{{\rm III}}+S^{{\rm IV}}-S^{{\rm V}}.\nonumber
\end{equation}
Here $S^{{\rm I}}-S^{{\rm V}}$ are pieces of action calculated at different
branches and are defined by:
\begin{eqnarray}
\begin{array}{l}
S^{\rm I}=S^{\rm IV}=\int_0^\infty d\chi\sqrt{2U(\chi)},\\
S^{\rm II}=S^{\rm III}=\int_{\chi_t}^\infty d\chi\sqrt{2(U(\chi)-\varepsilon)},\\
S^{\rm V}=\int_{-\tilde {\chi_t}}^{\tilde {\chi_t}} d\sqrt{2(U(\chi)-\varepsilon)}.
\end{array}
\end{eqnarray}
Clearly the branches $S^{{\rm I}},S^{{\rm IV}}$ correspond to
the part of trajectory with zero $\varepsilon$, while the branches $S^{%
{\rm II}}$,$S^{{\rm III}}$, $S^{{\rm V}}$- to those with nonzero $\varepsilon$.
The branch $S^{{\rm V}}$becomes zero, if the dimensionality is higher then
potential barrier. One can show, that each of $S^{{\rm I}}-S^{{\rm IV}}$
diverges, but the sum is finite.

4. THE MASSLESS SCALAR THEORY.

Consider the simplest example of scalar massless theory with the
Euclidean action
\begin{equation}
S=\int d^4x \Bigl\{\frac{1}{2}\partial_\mu\phi(x) \partial_\mu\phi(x)+\Bigr.
\Bigl.\frac{g^2}{4}{\phi^4}(x)\Bigr\}.
\end{equation}
In spherical coordinates it reads:
\begin{eqnarray}
S=\int_0^\infty d(\ln r)\Bigl\{{1 \over 2} ({\frac{\partial \chi(r,{\bf \alpha})}{\partial (\ln r)}})^2\Bigr.+
{1 \over 2} \chi (r,{\bf \alpha})(\hat L^2+1)\chi (r,{\bf \alpha})+
\Bigl.\frac{g^2}{4}\chi^4(r,{\bf \alpha})\Bigr\}
\end{eqnarray}
with the effective potential
\begin{eqnarray}
U_{eff}(\chi)={1 \over 2 }\chi^2 (r,{\bf \alpha})+\frac{g^2}{4}\chi^4(r,{\bf \alpha})
\end{eqnarray}
We are interested in singular field configurations,
parametrized by real $\ln r$. Assuming the angular independence of field
we get the following equation of motion:
\begin{eqnarray}
\frac{d^2\chi}{d(\ln r)^2}-\chi-g^2\chi^3=0
\end{eqnarray}
or
\begin{eqnarray}
\dot \chi^2=\frac{g^2}{2}\chi^4+\chi^2-2\varepsilon
\end{eqnarray}
with $\varepsilon$ as constant of integration.
For branches I and IV ($\varepsilon=0$) we obtain:
\begin{eqnarray}
\chi^{\rm I}(\tau)=-\frac{\sqrt2}{g}\frac{1}{\sinh (\tau+\tau_0/2)},\quad
{\rm for}\quad \tau<0,\\
\chi^{\rm IV}(\tau)=\frac{\sqrt2}{g}\frac{1}{\sinh (\tau-\tau_0/2)},\quad
{\rm for}\quad \tau>0.
\end{eqnarray}
For nonzero $\varepsilon$ the solutions are expressed in terms of
Jacobian elliptic functions:
\begin{eqnarray}
\chi^{\rm II}(\tau)=- {1 \over g} \sqrt{\frac{2\sqrt{1+4g^2\varepsilon}}
{{\rm sn}^2[\root 4\of {1+4g^2\varepsilon}(\tau+\tau_0/2)]}-(1+\sqrt{1+4g^2\varepsilon)}}
{\rm for}\quad  -\tau_0/2<\tau<0,\\
\chi^{\rm III}(\tau)={1 \over g} \sqrt{\frac{2\sqrt{1+4g^2\varepsilon}}
{{\rm sn}^2[\root 4\of {1+4g^2\varepsilon}(\tau-\tau_0/2)]}-(1+\sqrt{1+4g^2\varepsilon)}}
{\rm for} \quad \tau_0/2>\tau>0,
\end{eqnarray}
We see,that the solutions $\chi^{\rm I}(\tau)$ and $\chi^{\rm II}(\tau)$ are
singular at the points $\tau=-\tau_0/2$ and $\chi^{\rm III}(\tau)$,
$\chi^{\rm IV}(\tau)$ at $\tau=\tau_0/2$  and for $\varepsilon=0$
solutions $\chi^{\rm II,III}(\tau)$ coincide with $\chi^{\rm I,IV}(\tau)$ . The
 elliptic functions depend on parameter $\varepsilon$ defined by
\begin{eqnarray}
k=\frac{\sqrt{1+\sqrt{1+4g^2\varepsilon}}}{\sqrt{2}\root 4 \of {1+4g^2\varepsilon}}.
\end{eqnarray}
Besides they are complex-valued doubly periodic functions. The
turning points are defined by requirement $d\chi(\tau)/d\tau=)$,  of
which  the solutions are:
\begin{eqnarray}
\begin{array}{c}
\chi^{1,2}_t=\pm\sqrt{-1+\sqrt{1+4g^1\varepsilon}},\\
\chi^{3,4}_t=0.
\end{array}
\end{eqnarray}
The last one coincides with the
vacuum. So we have only two finite turning points at real $\tau$. It means,
that the branch $S^{{\rm V}}$ becomes zero and there is no tunneling in
theory. Collecting all these results and inserting into (24)-(26) we obtain:
\begin{eqnarray}
\frac{d\ln\sigma(\varepsilon)}{d\varepsilon}=-{1 \over {\root 4 \of {1+4g^2\varepsilon}}}{\bf K}(k),
\end{eqnarray}
where ${\bf K}(k)$ is a complete elliptic function of first
kind. Using the expansion of $K(k)$ in two limiting cases $\varepsilon \ll 1$
and $\varepsilon \gg 1$ and integrating over $\varepsilon$ we get respectively:
\begin{eqnarray}
\ln\sigma(\varepsilon)=-{1 \over 2}\varepsilon(\ln {16 \over {\varepsilon g^2}}+
1)+{1 \over 16}\varepsilon^2g^2(3\ln{16 \over {\varepsilon g^2}}-{25 \over 2})+O(\varepsilon^3)
\quad {\rm for} \quad \varepsilon \ll 1,
\end{eqnarray}
and
\begin{eqnarray}
\ln\sigma(\varepsilon)=-{[\Gamma (1/4)]^2 \over {3{\root 4 \of 4\pi^2g^2}}}
\varepsilon^{3/4}+ O(\varepsilon^{1/4}) \quad {\rm for} \quad \varepsilon \gg 1.
\end{eqnarray}
We see, that the cross section decreases as a function of $\varepsilon$ and
reproduces the result of [13].

5. SU(2)  YANG-MILLS THEORY.

The next model we consider is a pure Yang-Mills theory, the Euclidean
action of which is
\begin{equation}
S={\frac{1 }{4}}\int dx^4 F_{\mu\nu}^aF_{\mu\nu}^a
\end{equation}
with
\begin{equation}
F_{\mu\nu}^a=\partial_\mu A_\nu^a-\partial_\nu
A_\mu^a+g\epsilon_{abc}A_\mu^b A_\nu^c.
\end{equation}
For zero energies the singular trajectories have been indicated by S.Yu.
Khlebnikov[11]. They are BPST [15] instanton solutions with some
modification. We derive below the singular trajectories for all values
of $\varepsilon$ as exact solutions of equation of motion, using the ansatz
of BPST:
\begin{eqnarray}
A_\mu^a= {1 \over g}\eta_{\mu\nu}^a n_\nu{{\phi (r)}\over r},
\end{eqnarray}
where $n_\nu$ is again a unit vector, parametrized by
spherical coordinates and $\eta_{\mu\nu}^a$ are quantities introduced by
t'Hooft[16]. On substituting this Ansatz into the action last one reduces to
\begin{eqnarray}
S={3\pi^2 \over g^2}\int d\tau\bigl\{{1 \over 2}({d\phi \over d\tau})^2+
{1 \over 2}(\phi^2-2\phi)^2\bigr\}
\end{eqnarray}
with the double-well potential:
\begin{eqnarray}
U(\phi)={1 \over 2}(\phi^2-2\phi)^2. \nonumber
\end{eqnarray}
which has two minima equal to zero at $\phi =0,2$ and one maximum equal to
1/2 at $\phi =1$. For $\phi(\tau)$ one obtains the  equation:
\begin{eqnarray}
{d\phi \over d\tau}^2=(\phi^2-2\phi)^2-2\varepsilon.
\end{eqnarray}
For $\varepsilon =0$ we obtain:
\begin{eqnarray}
\phi^{\rm I}(\tau)=1+\coth (\tau+\tau_0/2),\quad {\rm for}\quad \tau <0,\\
\phi^{\rm IV}(\tau)=1-\coth (\tau-\tau_0/2),\quad {\rm for}\quad \tau >0,
\end{eqnarray}
while for nonzero $\varepsilon$ one gets two sets of solutions:\\
a)for $2\varepsilon <1$:
\begin{eqnarray}
\phi^{\rm II}_0=1+{{\sqrt{1+\sqrt{2\varepsilon}}} \over {{\rm sn}[\sqrt{1+
\sqrt{2\varepsilon}}(\tau+\tau_0/2)]}}, \quad {\rm for} \quad \tau <0,  \\
\phi^{\rm III}_0=1-{{\sqrt{1+\sqrt{2\varepsilon}}} \over {{\rm sn}[\sqrt{1+
\sqrt{2\varepsilon}}(\tau-\tau_0/2)]}},\quad {\rm for} \quad \tau >0,
\end{eqnarray}b)
b)for $2\varepsilon >1$:
\begin{eqnarray}\phi^{\rm I}=1+\sqrt{1-\sqrt{1+2\varepsilon}-\frac{2\sqrt{2\varepsilon}}
{{\rm sn}^2{\sqrt{2\sqrt{2\varepsilon}}(\tau+\tau_0/2)}}},\quad {\rm for}\quad \tau <0 \\
\phi^{\rm IV}=1-\sqrt{1-\sqrt{1+2\varepsilon}-\frac{2\sqrt{2\varepsilon}}
{{\rm sn}^2{\sqrt{2\sqrt{2\varepsilon}}(\tau-\tau_0/2)}}},\quad {\rm for}\quad \tau >0
\end{eqnarray}
It is easy to verify, that (47) and (48) coincide with (45),
(46) for $2\varepsilon \rightarrow 0$. Besides for $2\varepsilon=1$
(47) and (48) reduces to (49) and (50) correspondingly. At the turning
points $d\phi(\tau)/d\tau=0$. The solutions of this are:\\ a)for $2\varepsilon
<1$:
\begin{eqnarray}
\phi^{\rm II}_1=1+\sqrt{1+\sqrt{2\varepsilon}}, \\
\phi^{\rm III}_2=1-\sqrt{1+\sqrt{2\varepsilon}}, \\
\phi^{\rm II}_3=1+\sqrt{1-\sqrt{2\varepsilon}}, \\
\phi^{\rm III}_4=1-\sqrt{1-\sqrt{2\varepsilon}};
\end{eqnarray}
b)for $2\varepsilon >1$:
\begin{eqnarray}
\phi^{\rm II}_1=1+\sqrt{1+\sqrt{2\varepsilon}}, \\
\phi^{\rm III}_2=1-\sqrt{1+\sqrt{2\varepsilon}}, \\
\phi^{\rm II}_3=\phi^{\rm III}_4=0;
\end{eqnarray}

We see from (55)-(57), that there is no tunneling in this case. This is not
surprising, since the trajectory goes over the barrier. More interesting is
the case $2\varepsilon <1$, when the trajectory penetrates the barrier and
the branch $S^{{\rm V}}$ contributes to the cross section. Taking derivative
of $\ln \sigma(\varepsilon)$ with respect $\varepsilon$ as in the
massless scalar theory we obtain:\\ a)for $2\varepsilon <1$:
\begin{eqnarray}
\frac{d\ln \sigma(\varepsilon)}{d\varepsilon}=\frac{3\pi^2}{2g^2\sqrt{1+\sqrt{2\varepsilon}}}
{\bf K}\left(\sqrt{\frac{1-\sqrt{2\varepsilon}}{1+\sqrt{2\varepsilon}}}\right);
\end{eqnarray}
b)for $2\varepsilon >1$:
\begin{eqnarray}
\frac{d\ln \sigma(\varepsilon)}{d\varepsilon}=-\frac{3\pi^2}{2g^2\sqrt{2\sqrt{2\varepsilon}}}
{\bf K}\left(\sqrt{\frac{\sqrt{2\varepsilon}-1}{2\sqrt{2\varepsilon}}}\right)
\end{eqnarray}
Using well known expansions of complete elliptic functions in
two limiting values of $\varepsilon$ - $2\varepsilon \ll 1$ and $2\varepsilon \gg 1$
- one can obtain after integrating over $\varepsilon$ the following
expressions for the total cross section;\\
a)for $2\varepsilon \ll 1$:
\begin{eqnarray}
\ln\sigma(\varepsilon)=\frac{3\pi^2}{g^2}\left\{{1 \over 4}\varepsilon\right.
\left(\ln {32 \over \varepsilon}+1\right) - {3 \over 128}\varepsilon^2
\left.\left(\ln {32 \over \varepsilon}-{3 \over 2}\right)\right\}+O(\varepsilon^3);
\end{eqnarray}
b)for $2\varepsilon \gg 1$:
\begin{eqnarray}
\ln\sigma(\varepsilon)=-\frac{3\pi^2[\Gamma (1/4)]^2}{3{\root 4\of 8}g^2}
\varepsilon^{3/4}+O(\varepsilon^{1/4}).
\end{eqnarray}
Analysis of the obtained results shows rising behavior of the total
cross section for $2\varepsilon \ll 1$, whereas it decreases for $2\varepsilon \gg 1$.
Evidently the cross section reaches its maximal value, which occurs at
the value of $2\varepsilon$, defined by the top of barrier - in particular at $%
\varepsilon=1/2$. All these are in agreement with those of[13]. Summarize
all these one should mention, that there are infinite many ways in field
theory to reach the singularity. In our field theoretical approach this
way is chosen naturally. S, the use of BPST's ansatz makes it possible
to look for singular classical configurations without any further approximation.

The more realistic theory is studied and is the subject of other paper.

the author is indebted to H.W.J.M\"uller-Kirsten for helpful discussions.
This work was started at the Kaiserslautern University being supported by
DFG. It is also supported  by INTAS 93-1630EXT Grant.

\bf References \\
\begin{enumerate}
\item  M.Matiss, Phys.Rep. {\bf 214}, 159, (1992).

\item  P.G.Tinyakov, Int. J. Mod. Phys., {\bf A8},(1993),1823.

\item  A.Ringvald, Nucl.Phys. {\bf B330},(1990),1.

\item  S.Yu.Khlebnikov, V.A.Rubakov, P.G.Tinyakov, Nucl.Phys.{\bf B367}%
,(1991),334.

\item  A.M.M\"uller, Nucl.Phys.{\bf B348},310,(1991),{\bf B353}%
,(1991),44.

\item  V.Zakharov, Nucl.Phys. {\bf B385},(1992),452.

\item  I.Balitski, R.Braun, Phys. Rev. {\bf D47},(1993),1879.

\item  L.D.Landau, E.M.Lifshitz, Quantum mechanics, (Pergamon press,
New-York,1965).

\item  S.V.Iordanskji, L.D.Pitaevskji, Zh.Eksp.Teor. Fiz. {\bf 76},
(1979), 769, \\ (Sov.Phys. JETP,49,(1979),386).

\item  S.Yu.Klebnikov, Phys,Lett.{\bf B282},(1992),59.

\item  M.Voloshin, Phys.Rev. {\bf D43},(1991),1726.

\item  D.Diakonov, V.Petrov, Phys.Rev.{\bf D50},1994,266.

\item  S.Fubini,A.J.Hanson, R Jackiw, Phys.Rev. {\bf D7},(1973),1732.

\item  L.N.Lipatov,Zh.Eksp.Toer.Fis.,{\bf 72}%
,(1977),411,(Sov.Phys.JETP,{\bf 45},(1977),216).

\item  A.A. Belavin , A.M. Polyakov, A.S. Schwarz, Y.S. Tyupkin,
Phis.Lett {\bf 59B},(1975),85.

\item G.t'Hooft, Phus.Rev. D14, (1976),3432.

\end{enumerate}
\end{document}